\documentclass[final]{svjour2}
\usepackage{rotating}
\usepackage{color}
\usepackage[normalem]{ulem}
\usepackage[colorlinks]{hyperref}
\hypersetup{colorlinks,citecolor=red,linkcolor=blue,urlcolor=blue}
\usepackage{mathptmx}
\usepackage[numbers, sort&compress]{natbib}
\usepackage{graphicx,amsmath,amssymb,color}
\makeatletter \journalname{Journal of Low Temperature Physics}

\begin{document}

\newcommand{\hdblarrow}{H\makebox[0.9ex][l]{$\downdownarrows$}-}

\title{Universal behavior of quantum spin liquid and
optical conductivity in the insulator herbertsmithite}

\author{V.R. Shaginyan\and A.Z. Msezane\and V.A. Stephanovich\and K.G. Popov
\and G.S. Japaridze}

\institute{V.R. Shaginyan \at Petersburg Nuclear Physics
Institute RAS;
\\Gatchina, 188300, Russia, \at CTSPS, Clark Atlanta University,
\\ Atlanta, Georgia 30314, USA\\
\email{vrshag@thd.pnpi.spb.ru}\\ A.Z. Msezane \at CTSPS, Clark
Atlanta University,
\\ Atlanta, Georgia 30314, USA
\\V.A. Stephanovich \at Institute of Physics, Opole University,
\\Oleska 48, 45-052, Opole, Poland\\ \email{stef@uni.opole.pl}
\\K.G.Popov \at Komi Science Center, Ural Division, RAS, 3a,
Chernova str. Syktyvkar, 167982, Russia
\\ G. S. Japaridze \at Clark Atlanta University, Atlanta, GA 30314, USA}

\maketitle

\begin{abstract}

We analyze optical conductivity with the goal to demonstrate
experimental manifestation of a new state of matter, the
so-called fermion condensate. Fermion condensates are realized
in quantum spin liquids, exhibiting typical behavior of heavy
fermion metals. Measurements of the low-frequency optical
conductivity collected on the geometrically frustrated insulator
herbertsmithite provide important experimental evidence of the
nature of its quantum spin liquid composed of spinons. To
analyze recent measurements of the herbertsmithite optical
conductivity at different temperatures, we employ a model of
strongly correlated quantum spin liquid located near the fermion
condensation phase transition.  Our theoretical analysis of the
optical conductivity allows us to expose the physical mechanism
of its temperature dependence. We also predict a dependence of
the optical conductivity on a magnetic field. We consider an
experimental manifestation (optical conductivity) of a new state
of matter (so-called fermion condensate) realized in quantum
spin liquids, for, in many ways, they exhibit typical behavior
of heavy-fermion metals. Measurements of the low-frequency
optical conductivity collected on the geometrically frustrated
insulator herbertsmithite produce important experimental
evidence of the nature of its quantum spin liquid composed of
spinons. To analyze recent measurements of the herbertsmithite
optical conductivity at different temperatures, we employ a
model of strongly correlated quantum spin liquid located near
the fermion condensation phase transition. Our theoretical
analysis of the optical conductivity allows us to reveal the
physical mechanism of its temperature dependence. We also
predict a dependence of the optical conductivity on a magnetic
field.


\end{abstract}

\section{Introduction}
The herbertsmithite $\rm ZnCu_3(OH)_6Cl_2$ has a two-dimensional
(2D) triangular lattice with the geometric frustration
prohibiting the formation of spin ordering even at the lowest
accessible temperatures $T$, and has been interpreted as a
$S=1/2$ kagome antiferromagnet
\cite{helt,herb,herb2,herb3,Han:2012}. Magnetic kagome planes
consisting of Cu$^{2+}$ ions having spins $S=1/2$ are separated
by a nonmagnetic Zn$^{2+}$ layers. Observations have found no
evidence of a long-range magnetic order or spin freezing down to
temperature of 50 mK, indicating that $\rm ZnCu_3(OH)_6Cl_2$ is
the most optimal compound found to hold the quantum spin liquid
(QSL) \cite{helt,herb2,herb3,Han:2012}. These results are
confirmed by theoretical considerations demonstrating that the
ground state of kagome antiferromagnet is a gapless strongly
correlated quantum spin liquid (SCQSL)
\cite{pr,shaginyan:2011,shaginyan:2012:A,shaginyan:2011:C,Normand}.
Therefore, the above insulator properties offer unique insight
into the physics of QSL. Indeed, measurements of the heat
capacity reveal a $T$-linear term indicating that the low-energy
excitation spectrum from the ground state is gapless
\cite{helt,herb2,herb3}. The excitation spectrum can be obtained
from the low-temperature measurements of the heat conductivity
$\kappa(T)$. For instance, at $T\to 0$ residual value in
$\kappa/T$ signals that the excitation spectrum is gapless. The
presence of the residual value would clearly confirm the
presence of QSL in $\rm ZnCu_3(OH)_6Cl_2$, while the $\kappa/T
\to 0$ behavior suggests that the low-energy excitation spectrum
could have a gap \cite{jltp:2017}. The heat conductivity is
formed primarily by both acoustic phonons and itinerant spinons,
while the latter form QSL. Since the phonon contribution is
negligible to the applied magnetic field $B$, the elementary
excitations of QSL can be further explored by the magnetic field
dependence of $k$. Since the variation of the thermal
conductivity, measured under the application of magnetic field,
probes elementary itinerant excitations and is insensitive to
phonon's contributions, which "contaminate" $\kappa$, we suggest
that measurements of $\kappa(B)$ should shed light on the nature
of the ground state of QSL of herbertsmithite \cite{jltp:2017}.

Measurements of the real part of the low-frequency optical
conductivity $\overline{\sigma}$ as a function of the
temperature $T$ and the applied of magnetic field $B$ collected
on insulators with geometrical frustration produce important
experimental results regarding the nature of quantum spin liquid
composed of spinons \cite{Pilon,Lee}. Therefore we face a
challenging problem of interpreting the experimental data
\cite{Pilon} in a consistent way, including the $T$-dependence
of the conductivity.

In this paper we show that SCQSL represents the new state of
matter, and employ the model of SCQSL to explain the observed
value of the optical conductivity $\overline{\sigma}$ and its
temperature dependence in herbertsmithite. We demonstrate that
SCQSL is located near the topological fermion condensation
quantum phase transition (FCQPT), and predict a magnetic field
$B$ dependence of $\overline{\sigma}$. Our calculations are in a
good agreement with experimental data.

\section{Strongly correlated quantum spin
liquids} \label{QSL}

Strongly correlated quantum spin liquids represent a special
case of ordinary QSL, being a quantum state of matter composed
of spinons - chargeless fermionic quasiparticles having spin
$1/2$ \cite{shaginyan:2011,shaginyan:2011:C,jltp:2017,book}. In
the case of an ideal two-dimensional lattice of insulating
compounds, SCQSL can emerge provided that the geometrical
frustration of the lattice leads to a dispersionless
topologically protected branch of the spectrum with the zero
excitation energy known as the flat band
\cite{ks91,green,vol,vol1,vol2}. Then, FCQPT can be considered
as quantum critical point of SCQSL, composed of chargeless heavy
spinons with $S=1/2$ and the effective mass $M^*$, occupying the
corresponding Fermi sphere with the Fermi momentum $p_F$.
Consequently, the properties of insulating compounds coincide
with those of heavy-fermion metals with one exception. Namely,
the typical insulating compound resists the flow of the electric
charge \cite{shaginyan:2011,shaginyan:2011:C,shaginyan:2012:A}.
As we are dealing with compounds having non-ideal triangular and
kagome lattices, we have to bear in mind that the magnetic
interactions, impurities and possible distortion of the lattices
can shift the SCQSL from the exact FCQPT, positioning it
somewhere near it. Therefore, the actual location of the SCQSL
with respect to FCQPT has to be established solely from
experimental data analysis.

The usual approach to describe the systems with itinerant
fermions is the famous Landau Fermi liquid (LFL) theory
\cite{land}. The LFL theory is based on the mapping of the
strongly interacting system of real electrons and nuclei in a
solid to that of a weakly interacting Fermi gas. This implies
that the elementary excitations behave as corresponding
quasiparticles, determining the physical properties of the
system at low temperatures. These quasiparticle excitations have
a certain effective mass $M^*$, which is almost independent of
temperature, pressure, and magnetic field strength being a
parameter of the theory \cite{land,lanl1,PinNoz}. The LFL theory
fails to explain the experimental results related to the
dependence of $M^*$ on the temperature $T$, magnetic field $B$,
pressure and other external parameters;  in the vicinity of the
FCQPT, deviations from the LFL behavior are observed \cite{pr}.
These so-called non-Fermi-liquid (NFL) anomalies are generated
by large value of the effective mass $M^*$ associated with
FCQPT.

To make our paper self-contained, we begin with a brief outline
of the physical mechanism, responsible for the dependence of the
Landau quasiparticle effective mass, $M^*(B,T)$, on magnetic
field and temperature (we recollect that in LFL theory the
effective mass does not strongly depend on external parameters).
The key point here is that the effective mass begins to depend
strongly on temperature $T$, magnetic field $B$ and other
external parameters such as the pressure $P$ near the above
FCQPT \cite{pr}. The main feature of our approach is based on
the existence of one more instability channel for the Landau
Fermi liquid (in addition to the well-known Pomeranchuk
instability channel, see e.g. Ref. \cite{lanl1}). Namely, under
some conditions (see Ref. \cite{book} for details) the effective
mass of a Landau quasiparticle may become infinite. In this
case, to avoid unphysical situations related to either effective
mass divergence or even its negativity, the system alters its
Fermi surface topology so that the effective mass acquires
temperature and magnetic field dependencies. To investigate the
low temperature transport properties, the scaling behavior, and
the effective mass $M^*(B,T)$ of SCQSL, we use the model of
homogeneous heavy fermion liquid. In that case, the model allows
avoidance of complications associated with the crystalline
anisotropy of solids \cite{pr}, and the Landau equation,
describing the effective mass $M^*$ of  a heavy fermion liquid,
reads \cite{land,pr}
\begin{eqnarray}
\nonumber \frac{1}{M^*(B,
T)}&=&\frac{1}{M}+\sum_{\sigma_1}\int\frac{{\bf p}_F{\bf
p}}{p_F^3}F
({\bf p_F},{\bf p}) \\
&\times&\frac{\partial n_{\sigma_1} ({\bf
p},T,B)}{\partial{p}}\frac{d{\bf p}}{(2\pi)^3}, \label{HC1}
\end{eqnarray}
where $M$ is the corresponding bare mass, $F({\bf p_F},{\bf p})$
is the Landau interaction function, which depends on Fermi
momentum $p_F$, momentum $p$, with $F({\bf p_F},{\bf p})$ is
phenomenological function, obtained from the condition of the
best fit to experiment, and $\sigma$ is the spin index. For
simplicity, we assume that the Landau interaction does not
depend on the spins. At finite temperatures, the distribution
function $n$ can be expressed as
\begin{equation}
n_{\sigma}({\bf p},T)=\left\{ 1+\exp
\left[\frac{(\varepsilon_{\sigma}({\bf
p},T)-\mu_{\sigma})}T\right]\right\} ^{-1},\label{HC2}
\end{equation}
where $\varepsilon_{\sigma}({\bf p},T)$ is the single-particle
spectrum. In our case, the chemical potential $\mu$ depends on
the spin due to Zeeman splitting $\mu_{\sigma}=\mu\pm \mu_BB$,
$\mu_B$ is the Bohr magneton.

In LFL theory, the single-particle spectrum is a variational
derivative of the system energy $E[n_{\sigma}({\bf p},T)]$ with
respect to occupation number $n$,
$$\varepsilon_{\sigma}({\bf p},T)=
\frac{\delta E[n({\bf p})]}{\delta n_{\sigma}}.$$ Choice of the
interaction function shape and parameters is dictated to by the
fact that the system has to be at FCQPT \cite{pr,ckz,epl}. Thus,
the sole role of the Landau interaction is to bring the system
to the FCQPT point, where the Fermi surface alters its topology
so that the effective mass acquires temperature and field
dependence \cite{pr,ckz}. The variational procedure, applied to
the functional $E[n_{\sigma}({\bf p},T)]$, gives the following
explicit form for $\varepsilon_\sigma({\bf p},T)$,
\begin{equation}\label{epta}
\frac{\partial\varepsilon_\sigma({\bf p},T)}{\partial{\bf p}}
=\frac{{\bf p}}{M}-\int \frac{\partial F({\bf p},{\bf
p}_1)}{\partial{\bf p}}n_{\sigma}({\bf
p}_1,T)\frac{d^3p_1}{(2\pi)^3},
\end{equation}

Equations \eqref{HC2} and \eqref{epta} constitute the closed set
for self-consistent determination of $\varepsilon_\sigma({\bf
p},T)$ and $n_{\sigma}({\bf p},T)$ and the effective mass,
$p_F/M^*=\partial\varepsilon(p)/\partial p|_{p=p_F}$. We
emphasize here that in our approach the temperature and the
magnetic field dependence of the effective mass is derived
solely from the  $T$ and the $B$ dependence of
$\varepsilon_\sigma({\bf p})$ and $n_{\sigma}({\bf p})$. At the
point of FCQPT, Eq. \eqref{HC1} can be solved analytically
\cite{pr,ckz}. At $B=0$, the effective mass strongly depends on
$T$, demonstrating the NFL behavior
\begin{equation}
M^*(T)\simeq a_TT^{-2/3}.\label{MTT}
\end{equation}
At finite $T$, the application of magnetic field $B$ drives the
system the to LFL region with
\begin{equation}
M^*(B)\simeq a_BB^{-2/3}.\label{MBB}
\end{equation}

A deeper insight into the behavior of $M^*(B,T)$ can be achieved
using some "internal" (or natural) scales. Namely, near FCQPT
the solution $M^*(B,T)$ of Eq. \eqref{HC1} reaches its maximal
value $M^*_M$ at certain temperature $T_{M}\propto B$ \cite{pr}.
Hence, it is convenient to introduce the internal scales $M^*_M$
and $T_{M}$ to measure the effective mass and temperature
respectively. In other words, we divide the mass and temperature
by $M^*_M$ and $T_{M}$. This generates the normalized effective
mass $M^*_N=M^*/M^*_M$ and the temperature $T_N=T/T_{M}$. Near
FCQPT the normalized solution of Eq. \eqref{HC1} $M^*_N(T_N)$
can be well approximated by a simple universal interpolating
function \cite{pr}. The interpolation occurs between the LFL and
NFL regimes and represents the universal scaling behavior of
$M^*_N$ \cite{pr}
\begin{equation}M^*_N(y)\approx c_0\frac{1+c_1y^2}{1+c_2y^{8/3}}.
\label{UN2}
\end{equation}
Here, $y=T_N=T/T_{M}$, $c_0=(1+c_2)/(1+c_1)$, and $c_1$, $c_2$
are fitting parameters. The magnetic field $B$ enters Eq.
\eqref{HC1} only in the combination $\mu_BB/T$, making
$T_{M}\sim \mu_BB$. It follows from Eq.~\eqref{UN2} that
\begin{equation}
\label{TMB} T_M\simeq a_1\mu_BB,
\end{equation}
where $a_1$ is a dimensionless factor, $\mu_B$ is the Bohr
magneton. Thus, in the presence of magnetic field the variable
$y$ becomes $y=T/T_{M}\sim T/\mu_BB$. Taking Eq. \eqref{TMB}
into account, we conclude that Eq. \eqref{UN2} describes the
scaling behavior of the effective mass as a function of $T$ and
$B$: The curves $M^*_{N}$ at different magnetic fields $B$ merge
into a single one in terms of the normalized variable $y=T/T_M$.
Since the variables $T$ and $B$ enter symmetrically, Eq.
\eqref{UN2} also describes the scaling behavior of
$M^*_{N}(B,T)$ as a function of $B$ at fixed $T$.

\begin{figure}[!ht]
\begin{center}
\includegraphics [width=1.0\textwidth]{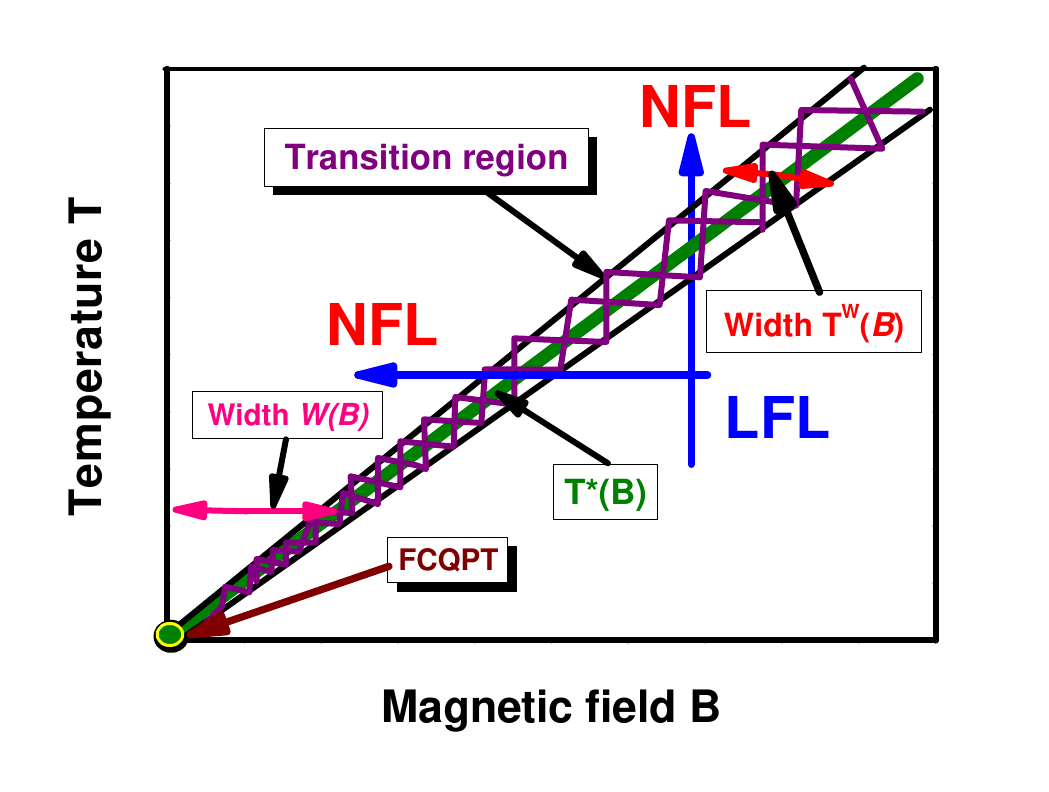}
\end{center}
\caption{(Color online). Schematic $T-B$ phase diagram of SCQSL
with magnetic field as the control parameter. The vertical and
horizontal arrows show LFL-NFL and NFL-LFL transitions at fixed
$B$ and $T$, respectively. The dependences of the effective mass
$M^*$ on $T$ and $B$ are given by Eqs. \eqref{MTT} and
\eqref{MBB}, respectively. The hatched area represents the
transition region taking place at $T^*(B)$, see Eq.~\eqref{BMT}.
The solid line in the hatched area represents the function
$T^*(B)\simeq T_M(B)$ given by Eq.~\eqref{TMB}. The functions
$W(B)\propto T\propto T^*$ and $T^W(B)\propto T\propto T^*$
shown by two-headed arrows define the width of the NFL state and
the transition area, respectively. At FCQPT indicated by the
arrow the effective mass $M^*$ diverges and both $W(B)$ and
$T^W(B)$ tend to zero.}\label{fig0}
\end{figure}
Now we construct the schematic phase diagram of SCQSL of the
insulator herbertsmithite. The phase diagram is reported in Fig.
\ref{fig0}. We assume for simplicity that at $T=0$ and $B=0$ the
system is approximately located at FCQPT without tuning. Both
magnetic field $B$ and temperature $T$ play the role of the
control parameters, shifting the system from FCQPT and driving
it from the NFL to LFL regions as shown by the vertical and
horizontal arrows. At fixed temperatures the increase of $B$
drives the system along the horizontal arrow from the NFL region
to LFL one. At the fixed magnetic field and increasing
temperatures the system transits along the vertical arrow from
the LFL region to the NFL one. The hatched area denoting the
transition region separates the NFL state from the weakly
polarized LFL one. The transition temperature $T^*(B)$ is given
by
\begin{equation}
\label{BMT} T^*(B)\simeq T_M(B),
\end{equation}
which directly follows from Eq. \eqref{TMB} ($a_2$ is a
dimensionless factor). The solid line represents the transition
region, $T^*(B)\simeq T_M(B)$. Referring to Eq.~\eqref{BMT},
this line is defined by the function $T^*\propto \mu_BB$, and
the width $W(B)$ of the NFL state is seen to be proportional to
$T$. In the same way, it can be shown that the width $T^W(B)$ of
the transition region is also proportional to temperature
\cite{pr}.

\section{Dynamic spin susceptibility and low-frequency optical conductivity}

\begin{figure} [! ht]
\begin{center}
\includegraphics [width=1.0\textwidth]{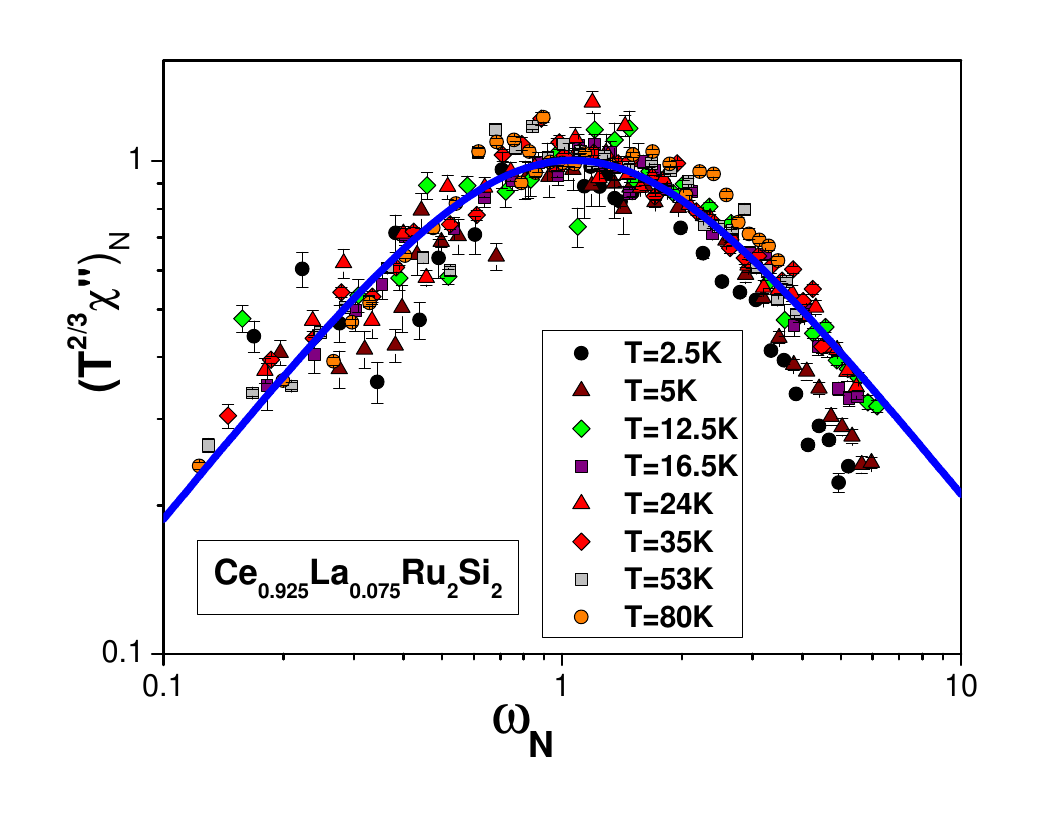}
\end{center}
\caption{(Color online) Scaling behavior of the normalized
dynamic spin susceptibility $(T^{2/3}\chi'')_N$. Data are
extracted from measurements on the heavy-fermion metal $\rm
Ce_{0.925}La_{0.075}Ru_2Si_2$ \cite{knafo:2004} and plotted
against the dimensionless variable $E_N$. Solid curve:
Theoretical calculations based on Eq.~\eqref{SCHIN}
\cite{shaginyan:2012:A}.}\label{fig04}
\end{figure}

\begin{figure} [! ht]
\begin{center}
\includegraphics [width=1.0\textwidth]{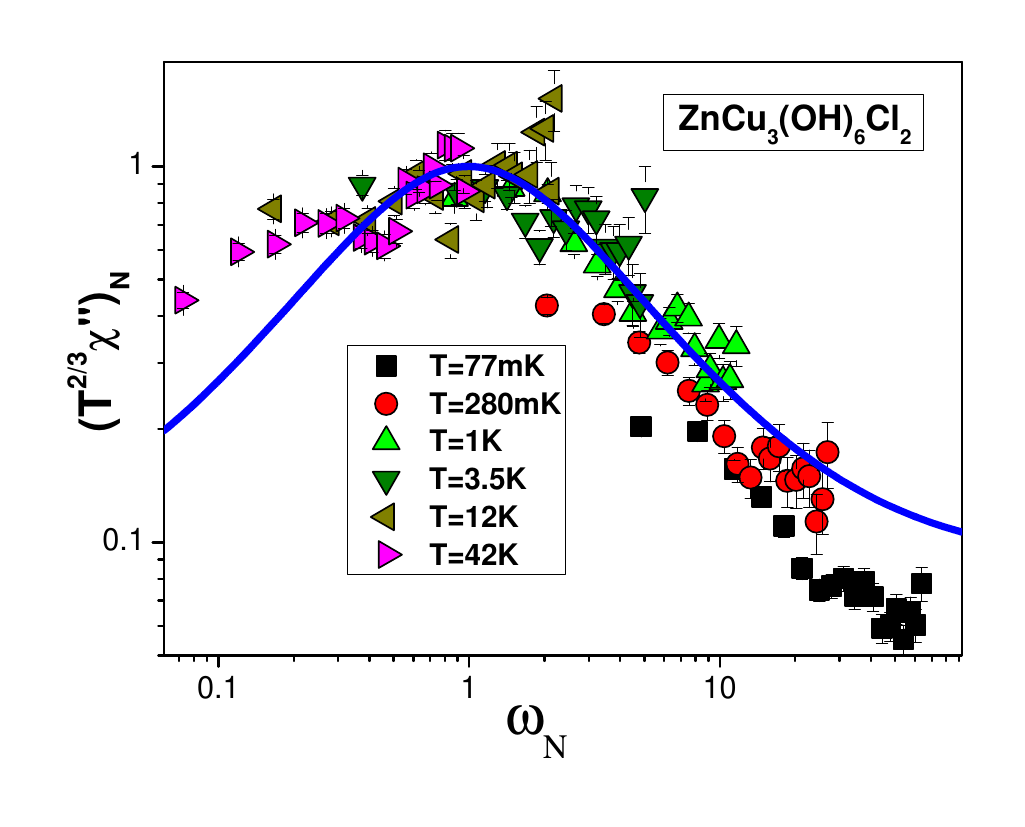}
\end{center}
\caption{(Color online) Scaling behavior of the normalized
dynamic spin susceptibility $(T^{2/3}\chi'')_N$. Data are
extracted from measurements on herbertsmithite $\rm
ZnCu_3(OH)_6Cl_2$ \cite{herb3}. Solid curve: Theoretical
calculations based on Eq.~\eqref{SCHIN}
\cite{shaginyan:2012:A}.}\label{fig05}
\end{figure}

Now we turn to the neutron-scattering measurements of the
dynamic spin susceptibility $\chi({\bf q},\omega,T)
=\chi{'}({\bf q},\omega,T)+i\chi{''}({\bf q},\omega,T)$ as a
function of momentum ${\bf q}$, frequency $\omega$, and
temperature $T$, former playing a crucial role in identifying
the properties of the quasiparticle excitations involved.  At
low temperatures, such measurements reveal that the
corresponding quasiparticles (belonging to herbertsmithite - a
new type of insulator) are represented by spinons, form a
continuum, and populate an approximately flat band crossing the
Fermi level \cite{Han:2012}. The imaginary part
$\chi''(T,\omega_1)$ satisfies the equation
\cite{shaginyan:2012:A,book}
\begin{equation}\label{SCHII}
T^{2/3}\chi''(T,\omega_1)\simeq\frac{a_1\omega_1}{1+a_2\omega_1^2},
\end{equation}
where $a_1$ and $a_2$ are constants and
$\omega_1=\omega/(T)^{2/3}$. It is seen from Eq. \eqref{SCHII}
that $T^{2/3}\chi''(T,\omega_1)$ has a maximum
$(T^{2/3}\chi''(T,\omega_1))_{\rm max}$ at some $\omega_{\rm
max}$ and depends on the only variable $\omega_1$. Equation
\eqref{SCHII} confirms the scaling behavior of $\chi'' T^{0.66}$
experimentally established in Ref. \cite{herb3}. As it was done
for the effective mass when constructing \eqref{UN2}, we
introduce the dimensionless function
$(T^{2/3}\chi'')_{N}=T^{2/3}\chi''/(T^{2/3}\chi'')_{\rm max}$
and the dimensionless variable $\omega_N=\omega_1/\omega_{\rm
max}$, and Eq. \eqref{SCHII} is modified to read
\begin{equation}\label{SCHIN}
(T^{2/3}\chi'')_N\simeq\frac{b_1\omega_N}{1+b_2\omega_N^2},
\end{equation}
with $b_1$ and $b_2$ are fitting parameters which are to adjust
the function on the right hand side of Eq. \eqref{SCHIN} to
reach its maximum value 1 at $\omega_n=1$. In such a situation
it is expected that the dimensionless normalized susceptibility
$(T^{2/3}\chi'')_{N}=T^{2/3}\chi'' /(T^{2/3}\chi'')_{\rm max}$
exhibits scaling as a function of the dimensionless energy
variable $\omega_N $ \cite{shaginyan:2012:A,book}. We predict
that if measurements of $\chi''$ are taken at fixed $T$ as a
function of $B$, then taking into account Eq. \eqref{MBB}, we
again obtain that the function $B^{2/3}\chi''(\omega)$ exhibits
the scaling behavior with $\omega_N=\omega_1/\omega_{max}$
\begin{equation}\label{SCHB}
(B^{2/3}\chi'')_N\simeq\frac{d_1\omega_N}{1+d_2\omega_N^2},
\end{equation}
In the same way, $d_1$ and $d_2$ are fitting parameters adjusted
such that the function $(B^{2/3}\chi'')_{N}$ reaches unity at
$\omega_N=1$. If the system is exactly at a FCQPT point, the
above scaling is valid down to lowest temperatures.
Figure~\ref{fig04} displays values of $(T^{2/3}\chi'')_{N}$
extracted from measurements of the inelastic neutron-scattering
spectrum on the heavy-fermion metal $\rm
Ce_{0.925}La_{0.075}Ru_2Si_2$ \cite{knafo:2004}. The scaled data
for this quantity, obtained from the measurements on another
quite different strongly correlated system such as $\rm
ZnCu_3(OH)_6Cl_2$ \cite{herb3}, are displayed in
Fig.~\ref{fig05}. It is seen that the theoretical results from
Ref.~\cite{shaginyan:2012:A} (solid curves) are in good
agreement with the experimental data collected on all two
compounds, for $(T^{2/3}\chi'')_{N}$ does exhibit the
anticipated scaling behavior for these systems. The scaled data
obtained in measurements on such quite different strongly
correlated systems as $\rm ZnCu_3(OH)_6Cl_2$ and $\rm
Ce_{0.925}La_{0.075}Ru_2Si_2$ collapse fairly well onto a single
curve over almost three decades of the scaled variables.

Now we are in a position to consider the low-frequency optical
conductivity $\overline{\sigma}$ of the insulator
herbertsmithite. We focus on low temperatures $T$ and
frequencies ${\omega}$, since the phonon absorption is expected
to contaminate the conductivity at elevated $T$ and $\omega$
\cite{Pilon}. In the atomic units $\hbar=c=1$, the Hamiltonian
of a particle with a spin reads
\begin{equation}\label{ham}
\hat{H}=\frac{1}{2m}\left({\bf p}-e{\bf
A}\right)^2+e\phi-\frac{\mu_B}{s}{\bf s}{\bf B},
\end{equation}
where ${\bf s}$ is the spin, ${\bf A}$ and ${\phi}$ are,
respectively, the vector and scalar potentials. Also, ${\bf
p}=i\nabla$ is the momentum operator of a particle, $e$ is its
charge. We use the gauge $\nabla{\bf A}=0$, so that ${\bf p}$
and ${\bf A}$ commute. In case of spinons $e=0$, and therefore
only the last term on the right hand side of Eq. \eqref{ham}
contributes to $\overline{\sigma}$.

As it follows from Eqs. \eqref{MTT} and \eqref{SCHII}, at low
frequencies $\omega$ the imaginary part of the spin
susceptibility is given by \cite{shaginyan:2012:A}
\begin{equation}\label{chi2}
\chi{''}(\omega)\propto \omega(M^*)^2.
\end{equation}
Taking into account that the energy transfer $\varepsilon_B$
\cite{PinNoz} from the magnetic field $B(\omega)$ to the system
is defined by the term $(\mu_B/{s}){\bf s}{\bf B}$ of Eq.
\eqref{ham}, we obtain
\begin{equation}\label{dEDtB}
\varepsilon_B=2\pi\omega\left[\frac{\mu_B}{s}{\bf s}{\bf
B}\right]\chi{''}(\omega) \equiv 2\pi \frac{\omega^2
\mu_B^3}{s}(M^*)^2{\bf s}{\bf B}.
\end{equation}
On the other hand, the energy transfer $\varepsilon_E$ from the
electric field $E(\omega)$ is given by
\begin{equation}\label{dEDtE}
\varepsilon_E=E^2(\omega)\overline{\sigma}(\omega).
\end{equation}
Comparison of Eqs. \eqref{dEDtB} and \eqref{dEDtE} yields
\begin{equation}\label{sigma}
\overline{\sigma}(\omega)\propto\omega\chi{''}(\omega)\propto
\omega^2(M^*)^2.
\end{equation}
From Eq. \eqref{sigma}  it follows that
$\overline{\sigma}(\omega)\propto \omega^2$. We note that
employed $\chi"$ given by Eq. \eqref{chi2} coincides at low
$\omega$ with that given by Eq. \eqref{SCHII}, and gives a
good description the facts, see see Figs. \ref{fig04} and
\ref{fig05}. At elevated temperatures it is seen from Eqs.
\eqref{MTT} and \eqref{sigma} the low-frequency optical
conductivity is a decreasing function of $T$. This
observation is consistent with the experimental data
\cite{Pilon}. It also follows from Eqs. \eqref{MBB} and
\eqref{sigma} that $\overline{\sigma}(\omega)$ diminishes
under the application of magnetic fields. Recent
observation seems to contradict the experimental results
since no systematic magnetic field dependence is observed
\cite{Pilon}. To elucidate the magnetic field dependence of
$\overline{\sigma}(B)$, we note that measurements of
$\overline{\sigma}(B)$ have been taken at 6 K and the
magnetic fields $B\leq 7$ T \cite{Pilon}. As it is seen
from Eq. \eqref{TMB} and Fig. \ref{fig0}, in such a case
the system is still in the transition regime and does not
enter into the LFL state at which the effective mass $M^*$
is given by Eq. \eqref{MBB}. Therefore, in this case the
effective mass behavior is determined by Eq. \eqref{MTT},
rather than by Eq. \eqref{MBB}, and the
$\overline{\sigma}(B)$ dependence cannot be observed. As a
result, we predict that the $B$-dependence of
$\overline{\sigma}$ can be observed at $B\simeq 7$ T
provided that $T\leq 1$ K.

\section{Conclusions}\label{SUM}

In summary, based on recent measurements of the low-frequency
optical conductivity in the herbertsmithite $\rm
ZnCu_3(OH)_6Cl_2$, we have shown that it can be viewed as a
strongly correlated Fermi system whose properties is defined by
SCQSL formed by chargeless spinons. Combining analytical
arguments and those based entirely on experimental data
analysis, we have come to conclusion that above quantum spin
liquid forms a new state of matter represented by fermion
condensate and discussed by us earlier (see, Refs.
\cite{pr,book,jltp:2017} for details). The crux of the matter
here is that geometrical frustration generates almost
dispersionless branches of the quasiparticles (spinons in this
case) spectrum (so-called flat bands), which makes the system
prone to ordinary LFL instability related to the quasiparticle
effective mass divergence \cite{pr,book}. This instability, in
turn, yields the fermion condensate - the new state of matter,
emerging during FCQPT. To analyze the experimental
manifestations of the above new state of matter in
herbertsmithite, we consider the dynamic spin susceptibility and
calculate the low-frequency optical conductivity of 2D
spin-carrying insulators with geometrical frustration. For that
we  employ the model of strongly correlated quantum spin liquid
located near the FCQPT. The comparison of our theoretical
results with experimental data on the herbertsmithite
low-frequency optical conductivity allows us to reveal its
temperature dependence. We also have predicted a dependence of
the optical conductivity on magnetic fields and pointed out
explicitly conditions at which such dependence can be
experimentally observed. Our findings not only establish the
existence of the robust SCQSL state in herbertsmithite, but also
provide a strategy for analyzing this state in insulators of new
type.

\begin{acknowledgements}
We are grateful to V.A. Khodel for valuable discussions.
This work was partly supported by U.S. DOE, Division of
Chemical Sciences, Office of Basic Energy Sciences, Office
of Energy Research.
\end{acknowledgements}


\end{document}